\documentclass{SciPost}

\hypersetup{
    colorlinks,
    linkcolor={red!50!black},
    citecolor={blue!50!black},
    urlcolor={blue!80!black}
}

\usepackage[bitstream-charter]{mathdesign}
\DeclareSymbolFont{usualmathcal}{OMS}{cmsy}{m}{n}
\DeclareSymbolFontAlphabet{\mathcal}{usualmathcal}

\fancypagestyle{SPstyle}{
\fancyhf{}
\lhead{\colorbox{scipostblue}{\bf \color{white} ~SciPost Physics Lecture Notes }}
\rhead{{\bf \color{scipostdeepblue} ~Submission }}

\fancyfoot[C]{\textbf{\thepage}}
}

\newcommand{\powerset}{\raisebox{.15\baselineskip}{\Large\ensuremath{\wp}}}

\usepackage{cleveref}

\begin{document}

\pagestyle{SPstyle}

\begin{center}{\Large \textbf{\color{scipostdeepblue}{
%%%%%%%%%% TODO: Write your article's title here
Complexity and accessibility of random landscapes
}}}\end{center}

\begin{center}\textbf{
% Use (full) first name (+ middle name initials) + surname format.
% Separate subsequent authors by a comma, omit comma and use "and" for the last author.
% Mark the corresponding author(s) with a superscript symbol in this order
% \star, \dagger, \ddagger, \circ, \S, \P, \parallel, ...
Sakshi Pahujani\textsuperscript{1$\star$} and 
Joachim Krug\textsuperscript{1$\dagger$}}
\end{center}

\begin{center}
{\bf 1} Institute for Biological Physics, University of Cologne,
K\"oln, Germany
%%%%%%%%%% END TODO: AFFILIATIONS
% Provide email address of corresponding author(s)
\\[\baselineskip]
$\star$ \href{mailto:spahuja1@uni-koeln.de}{\small spahuja1@uni-koeln.de}\,,\quad
$\dagger$ \href{mailto:jkrug@uni-koeln.de}{\small jkrug@uni-koeln.de}
\end{center}

\section*{\color{scipostdeepblue}{Abstract}}
\boldmath\textbf{%
  These notes introduce probabilistic landscape models defined on high-dimensional discrete sequence spaces. The models are motivated
  primarily by fitness landscapes in evolutionary biology, but links to statistical physics
  and computer science are mentioned where appropriate. Elementary and advanced results
  on the structure of landscapes are described with a focus on
  features that are relevant to evolutionary searches, such as the
  number of local maxima and the existence of fitness-monotonic paths. The recent discovery of submodularity as a biologically
  meaningful property of fitness landscapes and its consequences for their accessibility is discussed in detail.  
% The abstract is in boldface, and should fit in 8 lines. It should be written in a clear and accessible style, emphasizing the context, the problem(s) studied, the methods used, the results obtained, the conclusions reached, and the outlook. You can add a table contents, recommended if your paper is more than 6 pages long.
}

\vspace{\baselineskip}

%%%%%%%%%% BLOCK: Copyright information
% This block will be filled during the proof stage, and finilized just before publication.
% It exists here only as a placeholder, and should not be modified by authors.
\noindent\textcolor{white!90!black}{%
\fbox{\parbox{0.975\linewidth}{%
\textcolor{white!40!black}{\begin{tabular}{lr}%
  \begin{minipage}{0.6\textwidth}%
    {\small Copyright attribution to authors. \newline
    This work is a submission to SciPost Physics Lecture Notes. \newline
    License information to appear upon publication. \newline
    Publication information to appear upon publication.}
  \end{minipage} & \begin{minipage}{0.4\textwidth}
    {\small Received Date \newline Accepted Date \newline Published Date}%
  \end{minipage}
\end{tabular}}
}}
}
%%%%%%%%%% BLOCK: Copyright information

% For convenience during refereeing we turn on line numbers:
% \linenumbers

\vspace{10pt}
\noindent\rule{\textwidth}{1pt}
\tableofcontents
\noindent\rule{\textwidth}{1pt}
\vspace{10pt}

% Write your article contents here, starting from first \section.
% An example structure is given below.
\section{Introduction}
\label{sec:introduction}

Navigation on high dimensional landscapes is a theme that presents
itself in several fields. This is simply because real world systems
are complex many-body systems. A study of the dynamics of such systems
therefore involves functions of many variables, be it a free energy
function in disordered systems like spin glasses, cost functions in
optimisation problems defined on neural networks or fitness functions
quantifying the reproductive success of organisms
\cite{Ros_2022,Ros2025}. In such settings, the stationary points of these functions or landscapes are particularly of interest as they signify a potentially stable state of the system. Depending on how the landscape is constructed,  which in turn depends on the specific context, the landscapes may feature multiple stationary points. At first glance, one might infer that navigating to a globally stable 
point would be difficult owing to a tendency to get trapped in metastable states. Yet, as we will later explore in the text, this is not always true. Other properties that have piqued the interest of many are the dependence of accessibility of a stationary point on the initial state (i.e. the basin of attraction) and the distribution of the stationary points of a landscape on its domain.

Naturally over the years many advances have been made towards
characterising the properties of such stationary points. In this note
we deliver an account of some of these properties in the context of
biological evolution towards maxima of the fitness function, hereafter
referred to as peaks, driven by natural selection. As we will shortly
see, such a fitness function is defined on a combinatorially large
space. Two approaches to define a fitness function on this space are
discussed. The first approach assigns fitness values drawn randomly
from a distribution to elements of the domain. In this framework, the
statistics of the number of peaks are studied and their accessibility
characterised. Then we do the same for structured fitness landscapes
where fitness values are correlated to some extent. An important
  motivation for the study of structured fitness landscapes is
  the observation that empirical biological fitness landscapes display widely
  different degrees of ruggedness, and therefore cannot generally be
  well represented by an uncorrelated random model
  \cite{Szendro2013,deVisser2014,Bank2022}. We will return to this point in
  Section \ref{sec:wright_vs_fisher}. 

The models and methods used to arrive at the results presented here
are often borrowed from, closely associated with or may also
potentially inspire techniques to study high dimensional landscapes in
other contexts. For example, the House of Cards model discussed in
\cref{sec:hoc} is analogous to the exactly solvable Random Energy
Model \cite{Derrida_1980} of spin glasses. Another such connection is
stated in the form of a map from a structured fitness landscape --
Fisher's geometric model \cite{fisher,Hwang_2017} -- to the
antiferromagnetic Hopfield model \cite{Park_2020}, see Sect.~\ref{sec:construction}
for details. So while the mathematical structures employed to study fitness landscapes are enticing in their own right, these associations affirm their relevance beyond the confines of this text. 

Finally, a small disclaimer-- as in the short lectures on which this text is based, expressions for certain results in the following sections are taken from the literature without providing proofs. While readers are encouraged to approach these results as gateways to deeper inquiry, appreciating their role within the broader context of this note will suffice for the most part.

\section{Definitions}
\label{sec:definitions}

We begin by establishing the necessary definitions:

\begin{itemize}
	\item \textbf{Genotype.} A genotype is the hereditary information passed on from a parent to its offspring. Such information is carried by the nucleic acid polymers DNA and/or RNA. Here, we represent a genotype by a sequence $\sigma$ of length $L$,
	\begin{equation}
		\sigma = (\sigma_1, \sigma_2, \sigma_3, ..., \sigma_L),
	\end{equation}
	where the element $\sigma_i$ at a \textit{site} $i$ of the sequence is taken from an alphabet of size $a$ represented in this note by ${0,1,2,...,a-1}$. Each element of the alphabet set is called an \textbf{allele}. The sequence space is therefore the $L$-fold Cartesian product of the set of alleles,  
	\begin{equation}
		\sigma \in \{ 0,..., a-1 \}^L.
	\end{equation}

	\item \textbf{Hamming distance.} For a given pair of sequences $\sigma$ and $\tau$, the Hamming distance $d(\sigma, \tau)$ is defined as the number of differing sites between the two,
	\begin{equation}
		d(\sigma, \tau) = \sum_{i=1}^{L} (1-\delta_{\sigma_i, \tau_i}).
	\end{equation}
	\item \textbf{Hamming graph.} A set of sequences of length $L$, $\{ 0, ..., a-1\}^L$ equipped with the Hamming distance measure is a Hamming graph, $\mathbb{H}_a^L$. The binary Hamming graph, $\mathbb{H}_2^L$ is a hypercube in $L$ dimensions. \Cref{fig:binary_hamming} depicts such hypercubes for $L = 1, 2$ and $3$.
	
%figure1	
\begin{figure}[h!]
	\centering
	\includegraphics[width=0.9\textwidth]{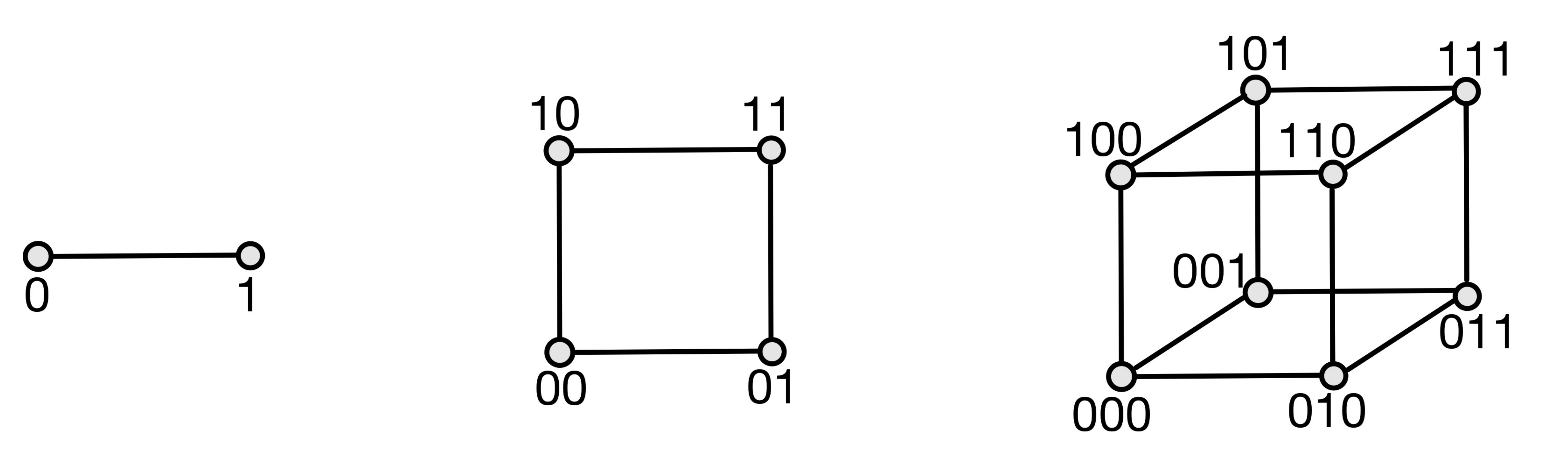}
	\caption{\textbf{Binary Hamming graphs.} From left to right, the figure shows hypercubes of dimension $L=1, 2$ and $3$. } 
	\label{fig:binary_hamming}
\end{figure} 

		\item \textbf{Mutations.} In general mutations are any form of changes in the genotype. In this note, we will consider only \textit{point mutations} which simply change the allele at one site of a genotype $\sigma$: 
	\begin{equation}
		\sigma_i \to \sigma_i^\prime.
	\end{equation}
      \item \textbf{Allele graph.} As such, Hamming graphs are constructed with an underlying assumption that any allele can mutate into any other allele from the allelic set. However, for a genetic sequence, this is not always true. To describe situations where the possible mutational transitions are constrained, we introduce the allele graph $\mathcal{A}$ \cite{Schmiegelt_2023}.
        This graph is defined over the allelic set and has an adjacency matrix $A = \{A_{\mu \nu}\}_{\mu,\nu=0,...,a-1}$ with
	
	\begin{equation}
	A_{\mu \nu} = 
		\begin{cases}
			1 \text{ if the transition from $\mu$ to $\nu$ is possible}\\
			0 \text{ otherwise}. 
		\end{cases}
	\end{equation}
	
	The sequence space can then be written as the $L$-fold Cartesian product $\mathcal{A}^L$ and the Hamming graph $\mathbb{H}_a^L$ is the special case when the allele graph $\mathcal{A}$ is complete. Some examples of allele graphs are depicted in \cref{fig:allele_graphs}.
	
%figure2
\begin{figure}[h!]
	\centering
	\includegraphics[width=\textwidth]{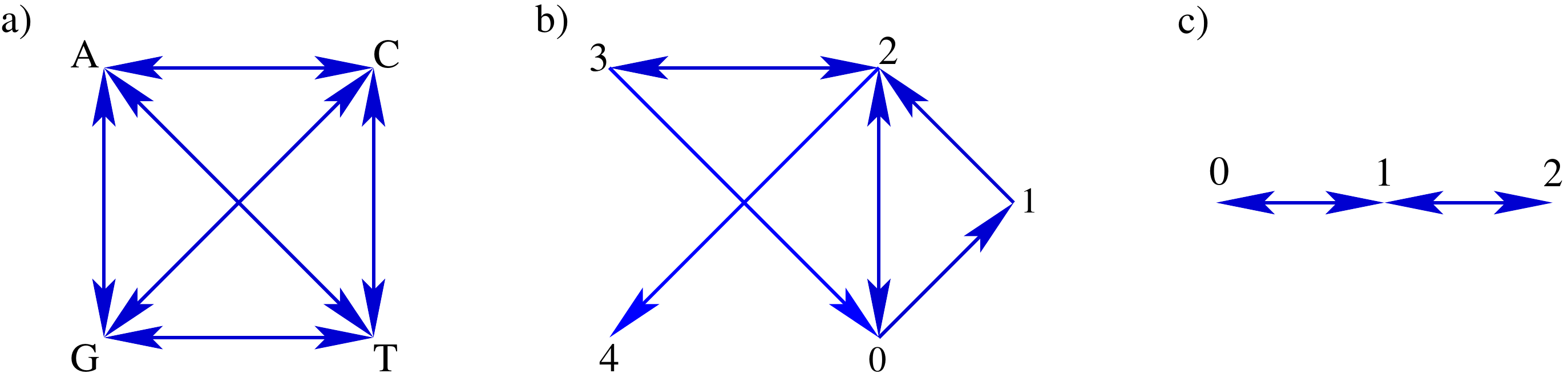}
	\caption{\textbf{Examples of allele graphs.} (a) The allele
          graph of nucleotides (constituents of DNA). (b) An
          incomplete allele graph over five
          alleles. This is an example of an allele graph for which the
          bound $\beta^\ast$ on the accessibility threshold $\beta_c$ discussed
          in Sect.~\ref{sect:indirect} is not
          tight, i.e., $\beta_c > \beta^\ast$. (c) An allele graph over three alleles (linear path
          graph). For this graph $\beta^\ast > 1$, which implies that
          there are no (direct or indirect) accessible paths for any
          fitness difference between initial and endpoint
          genotype. See \cite{Schmiegelt_2023} for details.} 
	\label{fig:allele_graphs}
\end{figure}

      \item \textbf{Fitness landscape.} From an evolutionary perspective, \textbf{fitness} is a measure of the reproductive capacity of an organism and is typically quantified by the mean number of offspring. A fitness landscape is a map $g : \mathcal{A}^L \to \mathbb{R}$ that
        assigns a fitness value to each genotype sequence. In the following we will assume that $g$ is non-degenerate, i.e.,
        no two genotypes have the same fitness value. 

      \item \textbf{Fitness graph.} By orienting the links between
        mutational neighbors in the direction of increasing fitness, a
        fitness landscape $g$ naturally induces an orientation on the
        genotype graph $\mathcal{A}^L$ which by construction is
        acyclic. The resulting acyclic oriented graph is referred to as a
        fitness graph \cite{deVisser2009,Crona2013,Crona2017,Riehl2022,Crona_2023}. Given a fitness graph, one is typically interested in the following two features:
	
	\begin{enumerate}
		\item \textbf{Peaks.} These are local or global maxima
                  of the fitness landscape. Peaks are the culminating
                  points (targets) of natural selection and appear as
                  sinks in the fitness graph. 
		\item \textbf{Accessible paths.} These are paths
                  (sequences of connected genotypes) on the fitness
                  landscape with monotonically increasing fitness. In
                  other words, an accessible path respects the orientation of
                  the fitness graph. This definition of accessibility
                    is similar to the concept of reachability introduced
                    in \cite{Stadler2010}.
	\end{enumerate}

\end{itemize}

\section{Wright's rugged vs. Fisher's accessible landscapes}
\label{sec:wright_vs_fisher}

The study of fitness landscapes dates back to the first half of the $20^{th}$
century. Since the beginning, these studies have been motivated by a
quest to understand the predictability and repeatability of evolution
\cite{deVisser2014}. At the core of this quest, one inevitably finds a
debate of ruggedness vs. accessibility.  Sewall Wright, who introduced
fitness landscapes in 1932, was of the opinion that
these are highly \textit{rugged} which implies that it is rather
difficult for an evolving population to reach the global fitness
maximum because of the many local maxima present in the
landscape \cite{Wright_1932}. Interestingly, Ronald Fisher, another foundational personality in the field
of evolutionary theory, had a contradictory opinion \cite{Provine1986}. He argued that the higher the number of dimensions, the greater the number of inequalities that must be satisfied  for a genotype to be a fitness maximum, as opposed to a single inequality in one dimension. Consequently, most genotypes would turn out to be saddles rather than peaks. 

To what extent real fitness landscapes are rugged or accessible is, at
heart, an empirical question which has begun to be explored over the
past two decades
\cite{deVisser2009,deVisser2014,Weinreich2006,Carneiro2010,Schenk2013,Aguilar2017,Pokusaeva2019,Bank2022}.
A fully consistent picture is still to emerge from this work. 
For example, a systematic study of 15 point mutations required to
mutate the spike protein responsible for cell entry from the Wuhan
strain of SARS-CoV2 to the Omicron BA.1 variant
found that the latter was not accessible via any of the $15!$ direct
paths\footnote{See Sect.~\ref{sec:accessibility} for a definition of this term.}, since
intermediate genotypes tended to lead to a lower affinity of the
protein to the cell wall receptor
\cite{Moulana_2022}. The interpretation offered by the authors of this
study is that the Omicron strain spread in the population not because
of higher transmissability, but because of its ability to overcome the
human immune defense; stated differently, receptor affinity is an
inapproprate fitness proxy in this case. 
On the other hand, a recent study inspecting nine nucleotide positions
of the DHFR gene in \textit{Escherichia coli} under antibiotic
selection pressure revealed that the resulting landscape is highly
rugged but also highly accessible \cite{Papkou_2023} (see also \cite{Westmann2024}). To eludicate the
relation between ruggedness and accessibility in fitness landscapes is a
major goal of the theoretical investigations that will be described in
these notes \cite{Krug_2024}. 

\section{House of Cards Model}
\label{sec:hoc}

\subsection{Ruggedness and complexity}

\label{sec:HoCcomplexity}

The House of Cards (HoC) landscape is a model where the fitness
values assigned to the elements of the sequence space, $g(\sigma)$, are
i.i.d. continuous random variables
\cite{Kingman_1978,Kauffman_1987}. We first try to gauge the level of
ruggedness present in this basic fitness model. The question we
therefore ask is the following: under this model, what is the probability that a certain genotype $\sigma$ is a peak?

To answer this, consider a set of genotypes in a sequence space $\mathbb{H}_a^L$ where $a$ and $L$ are as defined before. For a sequence to be an immediate neighbour of a given reference sequence in this set, it should have a Hamming distance of one from the latter. This means that a neighbour will have any one out of all $L$ positions differing from the reference and this differing position can have any one of the remaining $a-1$ values to choose from. So, a given sequence in this space will have $(a-1)L$ immediate neighbours. We pick a random sequence $\sigma^{(0)}$ with fitness $g_0$ which has neighbours $\{\sigma^{(i)}\}$ with corresponding fitnesses denoted by $\{g_{i}\}$ where $i \in \{ 1, ..., (a-1)L\}$.

We define $\mathcal{P}_{max}$ as the probability that a random
genotype $g_0$ is a peak, i.e., it has the highest fitness among all
its immediate neighbours. By symmetry \cite{Kauffman_1987}
\begin{align}
	\mathcal{P}_{max} = \text{Prob} \left[  g_0 = \text{max} \left\{ g_0, g_1, g_2, ..., g_{(a-1)L}  \right\}\right] 
	= \frac{1}{(a-1)L +1}. 
	\label{eq:p_max}
\end{align}
We next define the random variable $N_L$ as the number of peaks of the
landscape. The expected number of peaks is then given by
\begin{align}
	\mathbb{E} (N_L) = \mathcal{P}_{max} a^L = \frac{a^L}{(a-1)L
  +1}. 
	\label{eq:num_peaks}			    
\end{align}
In a sense, 
this simple result already suffices to resolve the controversy between
Sewall Wright and Ronald Fisher described in Section
\ref{sec:wright_vs_fisher} \cite{Franke2011}.
Equation (\ref{eq:p_max}) tells us that $\lim\limits_{L \to \infty}
\mathcal{P}_{max} = 0$ which supports Fisher's argument: A local maximum
is a rare occurrence in a high-dimensional landscape. On the other
hand, \eqref{eq:num_peaks} tells us that $\mathbb{E}(N_L) \to \infty$ for $L \to \infty$  in conjunction with Wright's argument that
such landscapes typically possess a large number of peaks.

One can also find the variance of the number of peaks to be \cite{Macken1989}

\begin{equation}
  \label{eq:var_peaks}
	Var(N_L) = \frac{a^L (a-1) (L-1)}{2 \left\{ (a-1)L +1 \right\}^2}.
\end{equation}
In the limit $L \to \infty$ 
\begin{equation}
	Var(N_L) \to \frac{1}{2} \frac{a^L}{(a-1)L} \approx \frac{1}{2} \mathbb{E}(N_L)
\end{equation}
which shows that the number of peaks has sub-Poissonian
statistics: Because two neighboring genotype sequences cannot both
be peaks, there is a local repulsion between peak genotypes which
reduces the fluctuations compared to a completely random distribution.  

The exponential dependence of \cref{eq:num_peaks} on $L$ suggests to
define the complexity $\Lambda$ of the landscape as the exponential
growth rate of the expected number of peaks with the dimensionality \cite{Ros_2022}
\begin{equation}
	\Lambda = \lim\limits_{L \to \infty} \frac{1}{L} \log \mathbb{E}(N_L).
	\label{eq:complexity}
\end{equation}
From \cref{eq:num_peaks}, $\Lambda = \ln a $, which means that to
exponential order the number of peaks is comparable to the number of
nodes in the graph, i.e., the landscape is (in this sense) maximally
rugged.

Note that in \eqref{eq:complexity} we have defined $\Lambda$ in terms
of the logarithm of the expectation
$\log \mathbb{E}(N_L)$. The resulting quantity is known as the
\textit{annealed} complexity \cite{Ros_2022}. For random landscapes in which
the number of peaks fluctuates strongly from one realization to the
other, the annealed complexity may not coincide with its
\textit{quenched} counterpart defined in terms of
$\mathbb{E}(\log N_L)$ \cite{Park_2020,Ros2023}. For the HoC model the two
definitions coincide, because [as can be read off from
\eqref{eq:num_peaks} and \eqref{eq:var_peaks}] the relative
fluctuations in $N_L$ vanish for $L \to \infty$ and the number of
peaks satisfies a central limit theorem \cite{Baldi1989}.

\subsection{Accessibility}

\label{sec:accessibility}

\begin{figure}[h!]
	\centering
	\includegraphics[width=0.9\textwidth]{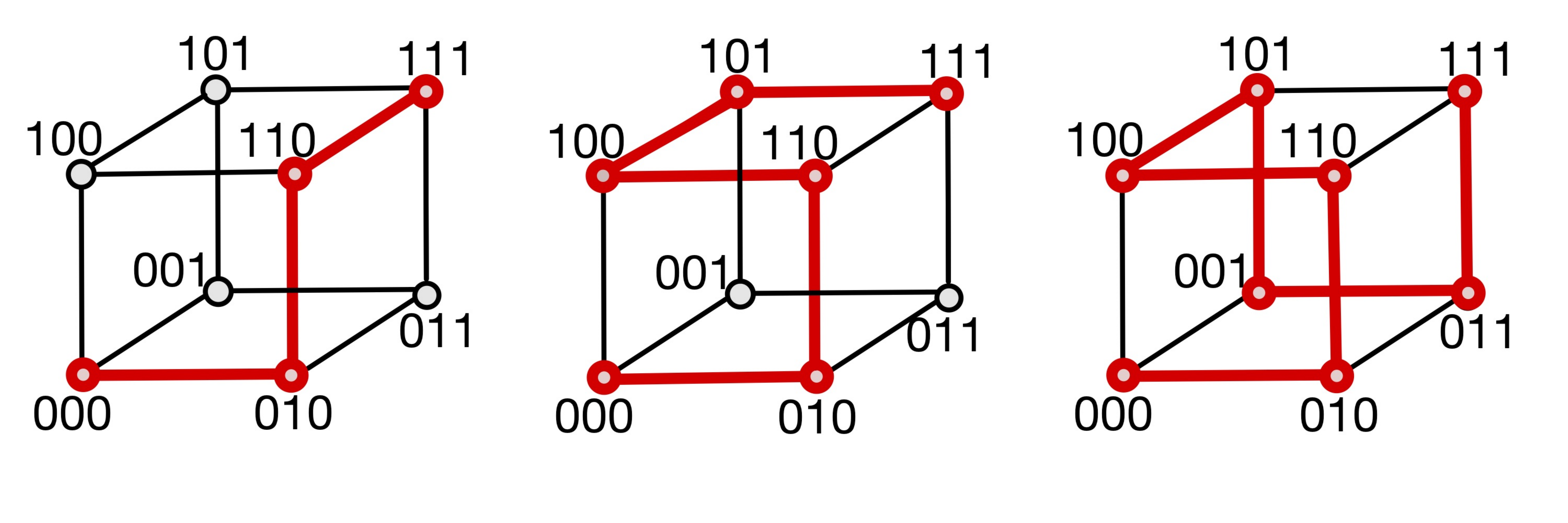}
	\caption{\textbf{Direct and indirect paths connecting the
            corners $\alpha=000$ and $\omega=111$ of the binary
            3-cube.} The figure shows a direct path of length $l =
          d(\alpha, \omega) = 3$ to the left and two indirect paths of
          length $l = d + 2 = 5$ and $l = d+ 4 = 7$
          next to it. The rightmost path visits all nodes and is
          the longest possible self-avoiding path.} 
	\label{fig:direct_indirect}
\end{figure}

Consider two genotypes $\alpha, \omega \in \mathcal{A}^L$ and a path
of length $l$ connecting them.
Let $\{\sigma^{(i)}\}$ be the set of sequences that this path goes through with the index $i$ representing the order in which they appear in the path with $i \in \{0, 1,\dots, l\}$. The path can be represented as

\begin{equation*}
 	\alpha = \sigma^{(0)} \to \sigma^{(1)} \to \sigma^{(2)} \to ... \to \sigma^{(l)} = \omega
\end{equation*}
such that any two consecutive sequences in the path are at a unit distance to each other,
\begin{equation}
	d(\sigma^{(i)}, \sigma^{(i+1)}) = 1. 
	\label{eq:unit_increase}
\end{equation}
The path is called \textit{accessible} if fitness always increases
monotonically along the path \cite{Weinreich2006,Carneiro2010,Franke2011,Weinreich2005}, 
\begin{equation}
	g(\sigma^{(i+1)}) > g(\sigma^{(i)}).
\end{equation}
It is \textit{direct} if the distance from the first sequence $\alpha$ in the path also increases monotonically along the path to the final sequence. Given \cref{eq:unit_increase}, this implies that the distance, of any sequence in a direct path, from the starting point is given by its position in the path,
\begin{align}
	&d(\alpha = \sigma^{(0)}, \sigma^{(i)}) = i.
	\label{eq:direct_path}
\end{align}
Otherwise, the path is \textit{indirect}. \Cref{fig:direct_indirect} shows examples of direct and indirect paths between two genotypes on a three dimensional binary Hamming graph. 

%figure3

In the following, we take a closer look at the problem of
accessibility in the HoC landscape. First, we restrict ourselves to
direct paths, and inspect the probability of existence of at least one
such path conditioned on the fitness gain between the initial and the
final sequences. We then expand our radar to include the indirect
paths and expound on how the resulting dramatic increase in
accessibility is tackled.

\subsubsection{Accessibility via direct paths}

 Note that the accessibility properties of the
HoC model, as well as the results on fitness maxima discussed in
Sect.~\ref{sec:HoCcomplexity}, rely only on the rank order of fitness values and are
therefore manifestly independent of the fitness distribution (provided
it is continuous). To simplify the following considerations we will nevertheless
specify the fitness values to be uniformly distributed on the interval
$[0,1]$. 

By definition \cref{eq:direct_path}, all direct paths between the
given genotypes $\alpha$ and $\omega$ have a length $l = d(\alpha,
\omega)$. This implies that the two sequences differ at $l$ sites and
$l!$ direct paths can be constructed between them.
Given a fitness difference of $\beta \in (0,1] $ between the two
sequences such that $g(\omega) = g(\alpha) + \beta$, a direct
accessible path would pass through $l-1$ intermediate sequences.
The fitness values of the intermediate sequences should lie in
$(g(\alpha), g(\alpha) + \beta)$ which is true with probability
$\beta^{(l-1)}$, and additionally they should be traversed 
in order of increasing fitness. There are $(l-1)!$ equiprobable
orderings of the $l-1$ i.i.d. random variables
$\{g(\omega^{(i)})\}_{i=1,\dots,l-1}$, and the probability that they
display the correct ordering is $1/(l-1)!$. Therefore, the probability
for a direct path between two sequences $\alpha$ and $\omega$ that are at distance $l$ and differ in fitness by $\beta$ to be accessible can be written as
\begin{equation}
	\mathcal{P}_{\beta, l} = \frac{\beta^{l-1}}{(l-1)!}.
	\label{eq:p_direct}
\end{equation}
With this knowledge at hand,
the expected
value of the number $X_{\alpha,\omega}$ of such paths is simply the product of the number of direct paths and \cref{eq:p_direct},
\begin{equation}
	\mathbb{E}(X_{\alpha, \omega}) = l! \frac{\beta^{l-1}}{(l-1)!} = l \beta^{l-1}.
	\label{eq:expected_direct_paths}
\end{equation}
In the limit $l \to \infty$, this quantity vanishes as long as $\beta<1$. Markov's inequality further bounds the probability of having at least one direct path by the mean number of direct paths as
\begin{equation}
	\mathcal{P}[X_{\alpha, \omega} \ge 1] = \sum_{k=1}^{\infty} \mathcal{P}[X_{\alpha, \omega} = k] \le \sum_{k=1}^{\infty} k \mathcal{P}[X_{\alpha, \omega} = k] = \mathbb{E}(X_{\alpha, \omega}) 
	\label{eq:markov}
\end{equation}
and therefore also the former quantity vanishes for sequences that are
very far apart on the Hamming graph unless the fitness difference
between them is exactly 1. When $\beta = 1$ the bound imposed by
\cref{eq:markov} on $\mathcal{P}[X_{\alpha, \omega} \ge 1]$ diverges
and is not informative.

A more careful analysis extending to the
second moment of $X_{\alpha, \omega}$ reveals a threshold based
increase in $\mathcal{P}[X_{\alpha, \omega} \ge1]$ with a threshold
fitness difference $\beta_c$ lying close to one. To be precise,
defining an $l$-dependent threshold $\beta_c(l)$ as 
\begin{equation}
	\beta_c (l) = 1 - \frac{\ln{l}}{l},
	\label{eq:beta_crit_direct}
\end{equation}
Hegarty and Martinsson \cite{Hegarty_2014} proved that the probability
of existence of an accessible paths displays a percolation-like phase
transition as a function of the fitness difference between the initial
and final genotype in the following sense:
\begin{equation}
	\lim_{l \to \infty} \mathcal{P}[X_{\alpha, \omega} \ge 1 | \beta = \beta_l] = 
	\begin{cases}
		0  &\text{ for } \beta_l < \beta_c(l) \\
		1  &\text{ for } \beta_l > \beta_c(l).
	\end{cases}
\end{equation} 

As we will see in the following, also in more general settings the onset of accessibility often
occurs when the expected number of accessible paths
$\mathbb{E}(X_{\alpha, \omega})$ is equal to unity. Correspondingly, one can arrive at \cref{eq:beta_crit_direct} by solving 
\begin{equation}
	\mathbb{E}[X_{\alpha, \omega} | \beta = \beta_c(l)] = 1
      \end{equation}
      for large $l$.

\subsubsection{Accessibility via indirect paths}
\label{sect:indirect}

Once the imposition to be direct is removed, accessibility increases
significantly. This may seem counter-intuitive considering that
indirect paths are longer and therefore have a lower probability to be
accessible. However, this is surpassed by the sheer increase in the
number of possible paths between two sequences on removal of this
constraint \cite{Depristo2007,Wu2016}.

Let us first consider the case of the binary hypercube (\cref{fig:direct_indirect}).
For two sequences at distance $d$, the length $l$ of an indirect path
satisfies $l =d + 2p$, where $p$ is the number of back-steps
i.e. reversals of a mutation ($1\to 0$). The total excess length is
$2p$, because each reversal has to be compensated by an additional
forward step ($0 \to 1$). Additionally, the path must satisfy the
fundamental constraint of acyclicity, i.e.
it must be self-avoiding. The longest self-avoiding path passes
through all nodes of the graph and has length $2^L-1$. For $a>2$ there are not only back-steps but also ``sideways'' steps
that lead to an allele that is present neither in the initial nor in the final sequence of the path \cite{Wu2016,Zagorski2016}.

Unlike in the analysis of direct paths, one cannot simply use the
knowledge of the number of indirect paths between $\alpha$ and
$\omega$ because estimating it becomes increasingly difficult for
larger $L$ \cite{Berestycki2017}. Schmiegelt and Krug \cite{Schmiegelt_2023} came up with a clever solution to this problem by introducing an extended fitness landscape where each node on a fitness graph (genotype) is assigned a countably infinite sequence of i.i.d. fitness values. A self-intersecting path on the original landscape can then be mapped to a self-avoiding path by invoking the $k^{th}$ subsequent fitness value from the sequence of fitnesses at the $k^{th}$ recurrence of a node in the path. If such a path on the extended landscape is accessible, the corresponding path on the original landscape is called `quasi-accessible'.  Schmiegelt and Krug \cite{Schmiegelt_2023} proved that the probability of having at least one quasi-accessible path is equal to that of having at least one accessible path. Therefore, the problem can be modified to look for quasi-accessibility and the difficulty associated with enumerating all self-avoiding paths is bypassed by expanding the analysis to all possible (generally intersecting) paths.

It is shown in \cite{Schmiegelt_2023} that the expected number of
quasi accessible paths $\mathbb{E}[\tilde{X}_{\alpha, \omega}]$
between genotypes $\alpha$ and $\omega$ with a fitness difference
$\beta$ depends exponentially on $L$ as 

\begin{equation}
  \label{eq:expectation}
	\mathbb{E}[\tilde{X}_{\alpha, \omega}] \sim \prod_{k,l =0}^{a-1} \left[ (e^{\beta A})_{k,l}\right]^{p_{k,l}L}.
\end{equation}

Here, $p_{k,l}$ is the fraction of sites where $\alpha_i = k$ and
$\omega_i = l$, and $A$ is the adjacency matrix of the allele
graph. The matrix $p_{k,l}$ encodes the scaled distance $\delta$
between the endpoints through
\begin{equation}
  \delta = \lim_{L \to \infty} \frac{1}{L} d(\alpha, \omega) =
  1 - \sum_{k=0}^{a-1} p_{kk}.
  \end{equation}

Guided by the analysis of the directed case in the preceding section,
we consider the condition $\lim_{L \to \infty}
\mathbb{E}[\tilde{X}_{\alpha, \omega}]=1$. By \cref{eq:expectation} this translates into

\begin{equation}
	\mathbb{E}[\tilde{X}_{\alpha, \omega}]^{\frac{1}{L}} = \prod_{k,l =0}^{a-1} \left[ (e^{\beta A})_{k,l}\right]^{p_{k,l}} = 1.
	\label{eq:thres_indirect}
      \end{equation}
      It is easy to see that the left hand side of this equation is an
      increasing function of $\beta$. Denoting the solution of
      \cref{eq:thres_indirect} by $\beta^\ast$, it then follows from
      Markov's inequality \eqref{eq:markov} that $\lim_{L \to \infty}
      \mathcal{P}[X_{\alpha, \omega} \ge 1] = 0$, i.e., accessible
      paths do not exist for $\beta < \beta^*$. This shows that $\beta^*$ provides a lower
      bound on the accessibility threshold $\beta_c$. In particular, since
      $\beta \in (0,1]$ by construction, there are no accessible paths
      at any fitness difference if $\beta^*>1$. Moreover, for a large
      class of allele graphs (including in particular the complete
      graph over $a$ nodes) it can be shown that the lower bound
      $\beta^\ast$ is tight and the accessibility threshold is given
      exactly by the solution of \cref{eq:thres_indirect} \cite{Schmiegelt_2023}. 
If the genotype space is the binary Hamming graph, the condition takes the simpler form \cite{Berestycki2017}
\begin{equation}
	\sinh(\beta_c)^\delta \cosh(\beta_c)^{1-\delta} = 1.
\end{equation}

\section{Structured Landscapes}
\label{sec:structured}

Though plenty of interesting properties underscore their importance as
an attractive play-field, completely random landscapes discussed so
far are not realistic because the idea that every single mutation
replaces an organisms fitness by an independent random variable is
bizarre \cite{Kingman_1978}. This motivates the study of structured
landscape models where fitness values of nearby sequences are correlated. 

Here we briefly mention some paradigmatic examples. 

\begin{itemize}
	\item \textbf{Kauffman's NK model}: The NK model
          \cite{Kauffman_1989,Weinberger_1991,Hwang_2018} imposes
          structure on the landscape by introducing interaction sets 
          $\mathcal{B}_i \subseteq \mathcal{L} \equiv \{1,\dots,L\}$,
          where $i = 1,\dots, b$ and the size of the subsets is
          $|\mathcal{B}_i| = k \le L$. The fitness effect of a
          mutation is affected by the loci in the
          subsets that it belongs to, but independent of other
          loci. This is achieved by writing the fitness of a sequence $\sigma$ as a sum of contributions 

          \begin{equation}
		g(\sigma) = \sum_{i=1}^b g_i(\sigma_{i,1}, \sigma_{i,2}, ..., \sigma_{i,k})
	\end{equation}
	where the index $i,j$ denotes the $j$'th element of the set
        $\mathcal{B}_i$ and $g_i$ is a function that assigns an
        independent random variable to each of its $a^k$
        elements. Thus, each $g_i$ is an independent HoC
        fitness landscape on the Hamming graph $\mathbb{H}_a^{k}$.
	
	The subsets $\mathcal{B}_i$ can be tailored to suit the
        underlying biological context. They can be systematically
        overlapping, disjoint or completely random. By tuning the
        parameter $k$ one can effectively control the degree of
        ruggedness in the landscape. For $k=1$ the alleles contribute
        independently to fitness and can be individually optimized,
        which means that the landscape is single-peaked. On the other
        hand, for $k=L$ the $g_i$ are random functions defined on the
        entire sequence space $\mathbb{H}_a^{L}$ and the HoC model is
        recovered.

        Mathematical results for the NK model have been obtained
          primarily in the case of binary alleles ($a=2$), and we
          provide a brief summary in the following. A key insight is
          that the landscape properties in the large $L$
          limit differ qualitatively depending on whether this limit is
        carried out at fixed $k$, or as a joint limit $k, L \to
        \infty$ at fixed ratio $k/L < 1$ \cite{Hwang_2018}. In the latter case
        NK landscapes are similar to HoC landscapes, in the sense that
      the exponential complexity \eqref{eq:complexity} takes on its maximal value
      $\Lambda = \ln 2$. For fixed $k$ the complexity is less than
      $\ln 2$ and generally depends on both $k$ and the choice of
      interaction sets. Remarkably, while fixed $k$ NK landscapes
      are less rugged than HoC landscapes, they are much less
      accessible: For most commoly used choices of interaction sets,
      the existence of accessible paths between genotypes at distance
      $d(\alpha,\omega) \sim L$ is exponentially unlikely for large
      $L$ \cite{Krug_2024,Hwang_2018}.
	
	\item \textbf{Rough Mount Fuji model}: The RMF model
          \cite{Aita_2000,Neidhart_2014} defines the fitness of a
          sequence as a function of the distance from a reference,
          $\sigma^\star$ (typically a high fitness genotype) with some linearly added stochasticity, 
	
	\begin{equation}
		g(\sigma) = -cd(\sigma, \sigma^\star) + \xi_\sigma
	\end{equation}
	where $c>0$ is a constant and $\xi_\sigma$ are i.i.d. continuous random variables.
	
	The degree of randomness in the landscape can be varied by
        modifying $c$. For $c=0$, the resulting landscape is random
        and for sufficiently large $c$ (larger than the variability of
        the $\xi_\sigma$) one gets a smooth (Mt. Fuji-like)
        landscape with a single global fitness peak. Explicit
          expressions for the expected number of peaks can be derived
          for specific choices of the distribution of $\xi_\sigma$
          \cite{Neidhart_2014}. Moreover, Hegarty and Martinsson have
          proved that direct accesible paths to the reference sequence
        $\sigma^\star$ exist with probability 1 for $L \to \infty$
        whenever $c > 0$ \cite{Hegarty_2014}.

	\item \textbf{Landscapes with an intermediate phenotype}:
          These landscapes are generated using a fitness function
          defined as a composition of a linear and a non-linear map on
          the sequence space. In the simplest case
          the linear part maps the sequence to a scalar real number
          $z(\sigma)$ 
          and the composition takes the form \cite{Manrubia2021}
	
	\begin{equation}
		g(\sigma) = \Phi\left[ z(\sigma) \right], \;\;\;
                z(\sigma) =  \sum_{i=1}^L \sum_{\mu = 0}^{a-1} a_{i,\mu} \delta_{\sigma_i,\mu} 
		\label{eq:fisher}
	\end{equation}
	
	where the $a_{i,\mu} \in \mathbb{R}$ are random coefficients
        representing the effect of allele $\mu$ at site $i$, and $\Phi$ is a non-linear function.
	
	One motivation for such a composition is to capture, to some
        extent, the complex biological interactions that determine the
        fitness of an individual. Organismal fitness is a consequence
        of the interaction of an organism with its
        environment. However, the genotype of an organism does not
        directly interact with the environment. Rather, 
        the genotype governs the physical characteristics of the organism,
        referred to collectively as its \textit{phenotype}, which then
        in turn interacts with the environment. The latter interaction
        determines the reproductive success of the organism which,
        through the heredity of the genetic information, modifies the
        genotypic composition of the population. A schematic of this feedback cycle is depicted in \cref{fig:schematic}.
	
	In \cref{eq:fisher}, the linear combination $z(\sigma)$
        represents the genotype-phenotype map and $\Phi$ represents
        the phenotype-fitness map. The scalar phenotype $z$ is maximized
        by the unique sequence $\sigma^\ast$ obtained by setting
        each $\sigma^\ast_i$ to the allele with the largest coefficient
        $a_{i,\mu}$. If the map $\Phi$ is monotonic, this property is
        inherited by the fitness landscape $g$ which is then
        single-peaked. Multi-peaked landscapes can emerge if $\Phi$ is
        non-monotonic, which implies that an intermediate value of the
        phenotype is favored by selection \cite{Srivastava2022}, or if
        the phenotype is multi-dimensional \cite{Manrubia2021,Schenk2013}.  

        The best known example of a composite
        genotype-phenotype-fitness map is Fisher's geometric model (FGM)
        that was introduced almost a century ago with the aim to
        elucidate the effect of organismal complexity on the adaptive
        process \cite{fisher}, see \cite{Tenaillon2014} and
        \cite{Hwang_2017,Park_2020} for a modern perspective. In the
        standard setting of FGM the phenotype is represented by a
        real valued vector $\vec{z} \in \mathbb{R}^n$, where the
        phenotypic dimension $n$ quantifies the organismal complexity,
        and the phenotype-fitness map $\Phi(\vec{z})$ is a radially
        symmetric function with a unique optimum at the origin (see
        Sect.~\ref{sec:submodular} for further discussion of this model).

%figure4
\begin{figure}[h!]
	\centering
	\includegraphics[width=0.6\textwidth]{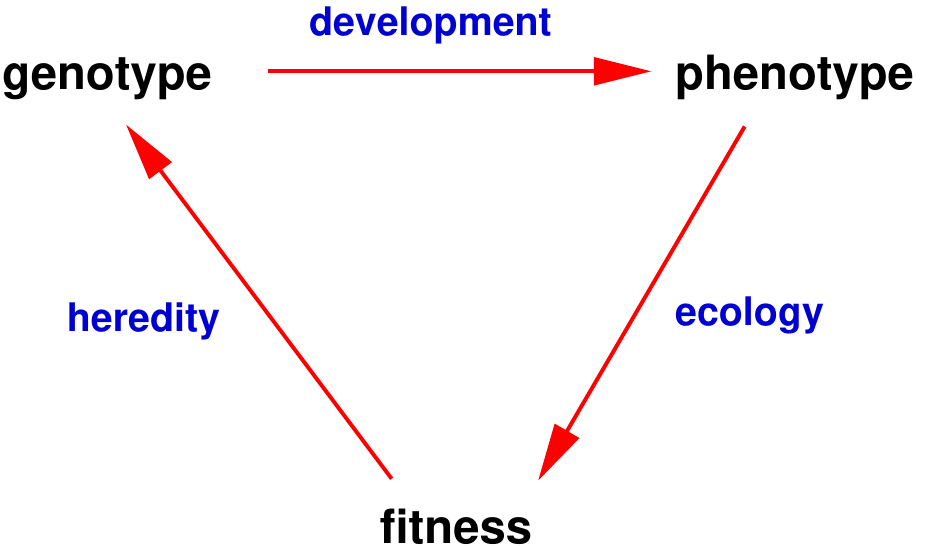}
	\caption{\textbf{Reductionist biology.} The feedback cycle of
          the key elements of Darwinian evolution. The genotype
          encodes all the ingredients for the development of an
          organism characterised by its phenotype (its physical
          features). Fitness, i.e. the reproductive success of the
          organism, is determined by its ecological interactions with
          the (biotic and abiotic) constituents of its
          environment. Based on its reproductive success, the relative
          abundance  of its genotype in the next generation is
          determined through the heredity of the genetic information. The
      figure was inspired by a lecture by Amitabh Joshi at JNCASR Bangalore.} 
	\label{fig:schematic}
\end{figure}

\end{itemize}

For the remaining sections, we will focus on composite landscapes, mostly with
a one-dimensional phenotype as defined in \cref{eq:fisher}. Moreover,
although many of the results that we will describe can be generalized
to the multiallelic setting, for convenience we consider only binary
genotype sequences
($a=2$). 

\section{Epistasis and submodularity}
\label{sec:epistasis}
\subsection{The concept of epistasis}

In classical genetics, 
epistasis is defined as the phenomenon that the effect of one gene variant
is cancelled by the presence of another
\cite{Phillips2008}. 
As a simple example, consider the effect of a gene variant causing
baldness on the gene responsible for hair colour. In the presence of
the former, the effect of the latter would be completely masked. In
modern usage, the term refers broadly to any kind of
interaction between mutational effects, both between and within genes
\cite{Bank2022,Weinreich2005,Poelwijk2016,Domingo2019}. A conceptual
difficulty with this definition is that, in order to identify interactions between
mutational effects, one needs to specify a non-interacting baseline
\cite{Krug2021a}.   

Consider first the simplest possible setting of binary sequences of
length $L=2$, and let the sequence $\sigma = 00$ represent the
reference genotype or `wild type'. With respect to it, the sequence
$11$ has two mutations and is therefore called `double mutant'. We denote by $s_{1/2}$ the fitness change in mutating the first(/second) site of the wild type, i.e. 
\begin{equation}
	s_{1} = g(10) - g(00),
\end{equation}
\begin{equation}
	s_2 = g(01) - g(00).
\end{equation}
If we assume no interactions between these two mutations, the expected
fitness of the double mutant, under an \textit{additive} null model, would be given by
\begin{equation}
	g_{add}(11) = g(00) + s_1 + s_2 = g(10) + g(01) - g(00).
\end{equation}
However due to epistasis, the actual fitness of the double mutant
might differ from this prediction. The pairwise epistatic interaction $\epsilon_{12}$ between the two mutants can therefore be quantified by the deviation from the additive expectation, 
\begin{equation}
	\epsilon_{12} = g(11) - g_{add}(11) = g(11)  + g(00) - g(10) -
        g(01). 
      \end{equation}
      
This reasoning can be generalized to larger $L$
\cite{Crona2017,Poelwijk2016,Neidhart2013,Faure2024}. For example, for $L=3$ there are three
pairwise epistatic interactions correponding to the double mutants
$110$, $101$ and $011$ and additionally a third order interaction
\begin{equation}
  \epsilon_{123} =
  g(000)+g(110)+g(101)+g(011)-g(100)-g(010)-g(001)-g(111).
\end{equation}
For binary genotypes of length $L$, there are $2^L-(L+1)$ epistatic
interactions. Together with the $L$ selection coefficients describing
the effects of single mutations and the wild type fitness
$g(0\dots0)$, the number of coefficients is thus equal to the number
of genotypes. This shows that the decomposition into epistatic
interactions retains the full information about the fitness
landscape. 

\subsection{Universal epistasis and submodularity}

\begin{figure}[h!]
	\centering
	\includegraphics[width=0.35\textwidth]{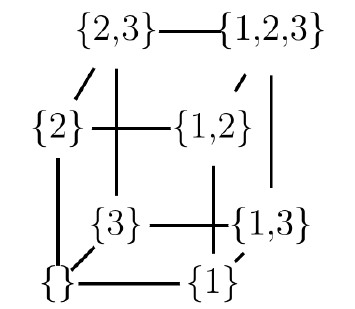}
	\caption{\textbf{Hasse diagram.} The diagram represents the
          nodes of the binary hypercube in three dimensions as
          elements of the power set $\mathcal{P}\{1, 2, 3\}$. Courtesy
        of Daniel Oros.} 
	\label{fig:hasse}
\end{figure}

For the purposes of the following considerations, it is useful to
write genotypes in set notation \cite{Crona_2023,Das_2020}. This is
done by noting that the binary hypercube $\{0,1\}^L$ is isomorphic to
the power set $\powerset ({\mathcal{L}}), {\mathcal{L}} = \{1,..,L\}$, such that for any $\sigma = (\sigma_1, \sigma_2, ..., \sigma_L) \in \{0,1\}^L$, 
\begin{equation}
	\sigma \rightarrow \{i: \sigma_i =1\} \in \powerset({\mathcal{L}}).
      \end{equation}
      This mapping is illustrated in Figure
      \ref{fig:hasse}. In the set representation $\sigma$
      comprises the subset of sites which have been mutated relative
      to the wild type sequence $0\dots0$.   

%figure5

Using the set representation, we can formalize a global constraint on
the fitness landscape that was first described in
\cite{Crona_2023}. Consider two genotypes $\sigma, \sigma'$, where
$\sigma'$ is a subset of $\sigma$, as well as a set $\tau$ of
mutations that are present in neither $\sigma$ nor $\sigma'$, i.e.,
$\tau \subseteq \mathcal{L} \backslash \sigma$. Then we say that the
fitness landscape displays \textit{universal epistasis} if the fitness
effect of adding the mutations in $\tau$
is always smaller (or always larger) in the genetic background of
$\sigma$ than in the background $\sigma'$. Specifically, the condition
of universal negative epistasis (UNE) reads \cite{Krug_2024}  
\begin{equation}
	g(\sigma \cup \tau) - g(\sigma) \le g(\sigma' \cup \tau) - g(\sigma').
	\label{eq:une}
\end{equation}
Universal epistasis implies that the
partial order induced by the subset-superset relation between genotypes is inherited by  
the fitness effect of mutations on the background of these genotypes.

\Cref{eq:une} turns out to be equivalent to a property known in discrete
mathematics as \textit{submodularity} for set functions
\cite{Krug_2024}. On defining $\sigma = A$ and $\sigma' \cup \tau =
B$, it follows that $\sigma \cup \tau = A \cup B$ and $\sigma' = A
\cap B$. Using these definitions and re-writing \cref{eq:une},
the inequality takes on the standard form of the condition for
submodularity \cite{Goldengorin2009,Krause2014}
\begin{equation}
	g(A \cup B) + g(A \cap B) \le g(A) + g(B) \text{  } \forall A,B \in \powerset(\mathcal{L}).
	\label{eq:submod}
\end{equation}
The equivalence of Eqs.~(\ref{eq:une}) and (\ref{eq:submod}) was first pointed out in \cite{Babayev1974}. Submodularity is a key
concept in discrete combinatorial optimization, because it is a property that arises naturally in many applications. The maximization
of submodular set functions is an NP-hard problem, but general theorems exist that provide bounds on the quality of
approximation algorithms, see e.g. \cite{Nemhauser1978}.

Conversely, one can define \textit{universal positive epistasis}
\cite{Crona_2023} by simply flipping the inequality in \cref{eq:une}
and show that it corresponds to the property of
\textit{supermodularity} written by similarly flipping the inequality
in \cref{eq:submod}. In the following, we will keep the
  discussion focused to UNE because of its important consequences for
  the accessibility of fitness peaks (see Sect.~\ref{sec:submodular}). By symmetry, universal positive
  epistasis and supermodularity have analogous consequences for the
  accessibility of the local minima of the fitness function via
  fitness-monotonic paths.

\subsection{Construction of submodular landscapes}
\label{sec:construction}

The concept of universal epistasis was originally formulated in the context of a geometric classification of fitness landscapes
that is based on the triangulation that a landscape induces in the \textit{continuous} cube $[0,1]^L$, see \cite{Beerenwinkel2007}
for details. Specifically, Crona et al. \cite{Crona_2023} showed that landscapes that induce a particular (staircase) triangulation satisfy the condition of universal positive epistasis. While they also reported a certain enrichment of this property in an empirical
data set, it is currently unknown to what extent biological fitness landscapes should be expected to satisfy conditions of
universal epistasis, sub- or supermodularity. Here we show, for a simple example, how submodularity may arise from a composite
genotype-phenotype-fitness map of the type discussed in Sect.~\ref{sec:structured}.

To see this, we first recall the defining property of a concave function $\Phi: \mathbb{R} \to \mathbb{R}$
\begin{equation}
  \label{eq:concave}
  \Phi(x + z) - \Phi(x) \leq \Phi(y + z) - \Phi(y) \;\; \forall x > y, \; z > 0
\end{equation}
and note its similarity to the UNE condition (\ref{eq:une}). Indeed, it is often stated that sub- and supermodularity can be regarded
as a generalization of concavity or convexity to set functions \cite{Krause2014,Babayev1974}. Consider now a fitness
landscape of the form \eqref{eq:fisher} where we assume binary genotypes ($a=2$) and moreover impose the condition
that the coefficients $a_i$ of the linear genotype-phenotype map are positive, $a_i > 0$. Then, by construction, the phenotype
\begin{equation}
	z(\sigma) = \sum_{i=1}^L a_i \sigma_i
\end{equation}
is a monotonically increasing function of the number of mutations in a genotype and therefore $z(\sigma')<z(\sigma)$ for any $\sigma' \subset \sigma$. If we further assume that $\Phi$ is concave, then
\begin{align}
	g(\sigma \cup \tau) - g(\sigma) &= \Phi[z(\sigma) + z(\tau)] - \Phi[z(\sigma)] \nonumber \\
	&\leq \Phi[z(\sigma') + z(\tau)] - \Phi[z(\sigma')] =
        g(\sigma' \cup \tau) - g(\sigma') 
	\label{eq:concavity}
\end{align}
where the inequality follows from the concavity property
(\ref{eq:concave}), and $z(\sigma \cup \tau) =
z(\sigma) + z(\tau)$ because $\sigma \cap \tau = \emptyset$.
This is precisely the condition (\ref{eq:une}) of submodularity or UNE.
With a little more work, one can verify that submodularity is obeyed
(in suitably transformed sequence space coordinates)  even when the
positivity condition on the $a_i$s is relaxed \cite{Krug_2024}.

\begin{figure}[h!]
	\centering
	\includegraphics[width=0.6\textwidth]{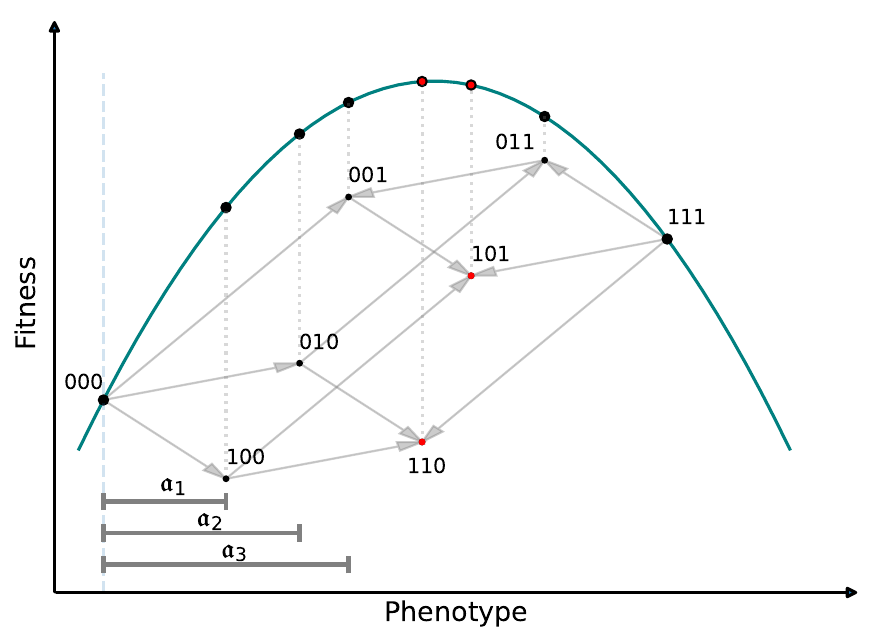}
	\caption{\textbf{Submodular landscapes construction.} A submodular landscape constructed by convolution of a linear genotype-phenotype map and a concave phenotype-fitness map. The three individual mutations increase the phenotypic value by different amounts, and the phenotypes of all other genotypes are linear combinations of these effects. Then, a concave function maps the phenotypes to their fitness values. The non-monotonicity of this function leads to a fitness graph with multiple peak (two in this case -- marked in red). The fitness graph in faint grey illustrates the rank ordering of the fitness values.} 
	\label{fig:sm_landscape}
\end{figure}

In the present context our main interest is in submodular fitness
landscapes that have multiple peaks. This can be achieved by choosing
$\Phi$ to be non-monotonic \cite{Manrubia2021}, see \cref{fig:sm_landscape} for illustration. With this
choice the model becomes a version of Fisher's geometric model (FGM), see Sect.~\ref{sec:structured}, and we conclude that the
fitness landscape of FGM with a one-dimensional phenotype is submodular (satisfies the condition of UNE) if
the phenotype-fitness map $\Phi$ is concave. 

%figure6

The problem of maximising fitness in fitness landscapes is analogous
to minimizing energy in a physical system, and correspondingly the
concept of submodularity can be applied to disordered
spin models as well. As an example, we choose the (concave)
phenotype-fitness map $\Phi = -z^2$ in Fisher's model and consider the
Hamiltonian
\begin{equation}
  H = -\Phi = z^2. 
\end{equation}
With the appropriate change of the $\sigma_i$'s to spin variables
$\eta_i = 1 - 2\sigma_i$, one arrives at the Hamiltonian of an
antiferromagnetic Hopfield model with continuous patterns and random fields \cite{Park_2020,Nokura_1998}, 
\begin{equation}
  \label{eq:Hopfield}
	H = \sum_{i,j} J_{ij}\eta_i \eta_j  +  \sum_{i} h_i \eta_i
 \end{equation}
 with 
 \begin{equation}
   \label{eq:J}
 	J_{ij} = \frac{1}{4} a_i a_j 
 \end{equation}
 and 
 \begin{equation}
   \label{eq:h}
 	h_i = -\frac{1}{2} \left( \sum_j a_j \right) a_i.
 \end{equation}
Based on the previous discussion, it follows that the Hamiltonian defined by Equations
(\ref{eq:Hopfield}-\ref{eq:h}) is a submodular function. Note that, by construction, the ground state 
of the model is $\eta_i \equiv 1$ ($\sigma_i \equiv 0$). 

\section{Accessibility of submodular landscapes}
\label{sec:submodular}

In this section, the reward for establishing the submodularity
property can finally be reaped with respect to one of the broader
themes of this text, i.e., the accessibility of local fitness peaks
via fitness-increasing paths. We already demonstrated that Fisher's
geometric model generates rugged submodular fitness landscapes when
certain conditions are met.
Here, we show that a specific kind of accessibility property is implicit to the construction of submodular landscapes.

\subsection{The accessibility property}

The aforementioned property is known by the name of
\textit{subset-superset accessibility property} (AP), and was first
identified for a landscape model with two intermediate phenotypes
designed to describe the evolution of antibiotic resistance
\cite{Das_2020,Das2022,Das2025}. We say that a landscape has the AP if \textbf{any
peak is accessible from all its sub- and supersets along all direct
paths}. It is trivial to see that any peak in such a landscape would
always be accessible from the zero-string i.e. $\sigma = \varnothing$
and the one-string $\sigma = \mathcal{L}$.

To prove that submodular landscapes have the AP, we consider a peak genotype $\sigma$. This means that any immediate neighbours in the subset or superset of $\sigma$ would have a fitness lower than $\sigma$ i.e.,
\begin{equation}
	g(\sigma \cup \{i\}) - g(\sigma) < 0
\end{equation}
and
\begin{equation}
	g(\sigma) - g(\sigma \backslash \{j\}) >0
\end{equation}
for all $j \in \sigma$, $i \in \mathcal{L} \backslash \sigma$.
Now consider a subset genotype $\sigma^\prime \subseteq \sigma$ and a
mutation $k \in \sigma \backslash \sigma' \subset \sigma$. Then by the
UNE condition \eqref{eq:une}, we have
\begin{equation}
	g(\sigma' \cup \{k\})  -  g(\sigma')  \geq  g(\sigma)  -  g(\sigma \backslash \{k\})  > 0.
\end{equation}
This implies that the mutation $k$ is fitness increasing on the background
$\sigma'$, ergo the corresponding step on the direct path from $\sigma'$ to $\sigma$ is accessible.
One can similarly prove the accessibility from a genotype in the superset of $\sigma$.

%figure7

\begin{figure}[h!]
	\centering
	\includegraphics[width=0.7\textwidth]{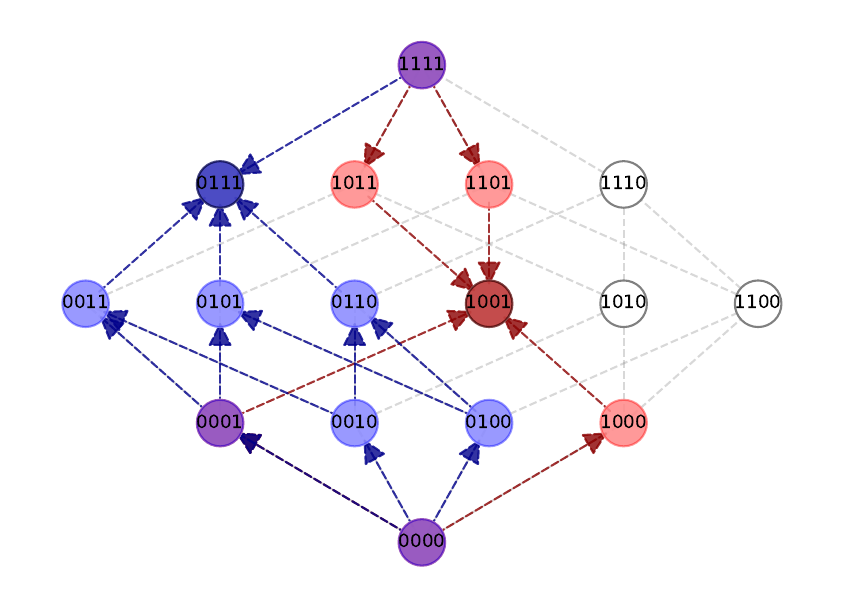}
	\caption{\textbf{Illustration of the subset-superset accessibility
        property.} Two peak genotypes (shown in dark red and dark blue) in the fitness graph are accessible from all their sub- and supersets (depicted in light red and light blue) as indicated by the arrows.} 
	\label{fig:sub_sup_ap}
\end{figure}

The proof outlined above was first presented in \cite{Krug_2024}, but
in fact the AP of submodular landscapes
has been known for a long time in the Russian literature on
combinatorial optimization, where it is attributed to a 1962 article
by Viktor Pavlovich Cherenin \cite{Goldengorin2009,Babayev1974}. It
was rediscovered in 1974 by Frieze \cite{Frieze1974}.

Importantly, while submodularity implies the AP, the two properties are not equivalent. Indeed, the AP uses only the rank
ordering of fitness values and, as such, it is invariant under
arbitrary monotonic transformations of the fitness values. This is in
contrast to the submodularity conditions
(\ref{eq:une},\ref{eq:submod}) which depend on the actual values
taken by the function $g$. As an application of this observation, consider the example of
FGM with a one-dimensional phenotype discussed in
Sect.~\ref{sec:construction}. Whereas submodularity is guaranteed for this model only when 
the phenoype-fitness map $\Phi$ is concave, the AP holds much more
generally, for any function $\Phi$ that has at most one maximum, that
is, for which the derivative changes sign at most once. This includes
in particular the Gaussian fitness function $\Phi(z) = e^{-z^2}$
that is often assumed in the context of FGM
\cite{Hwang_2017,Park_2020,Tenaillon2014}, as well as more generally in
the description of the evolution of phenotypic traits \cite{Burger2000,Pahujani2025}.

Exploiting the fact that a positive linear combination of
submodular functions is also submodular \cite{Krause2014}, we can
further establish the AP for a special case of FGM with a
multidimensional phenotype. Consider the phenotype-fitness map
\begin{equation}
  \label{eq:FGMmulti}
  \Phi(\vec{z}) = \exp\left[-\vert \vec{z} \vert^2\right] =
  \prod_{k=1}^n e^{-z_k^2},
  \end{equation}
and let the components $z_k$ of the $n$-dimensional phenotype vector
$\vec{z}$ depend on the genotype through the linear relation
\begin{equation}
  \label{eq:FGMpheno}
  z_k(\sigma) = q_k + \sum_{i=1}^L a_{i,k} \sigma_i, \;\;\; k = 1,
  \dots, n
\end{equation}
where the $q_k$ are arbitrary constants and the coefficients $a_{i,k}
> 0$. Then the logarithm of the fitness function $g(\sigma) =
\Phi[\vec{z}(\sigma)]$
\begin{equation}
  \label{eq:FGMlog}
  \ln g(\sigma) = \sum_{k=1}^n - \left(q_k +  \sum_{i=1}^L a_{i,k}
    \sigma_i \right)^2
\end{equation}
is a sum of concave functions of a linear genotype-phenotype map with
positive coefficients. On
the basis of the arguments of Sect.~\ref{sec:construction} we conclude
that $\ln g$ is submodular, and therefore $g$ has the AP. Note that,
in contrast to the one-dimensional case discussed in
Sect.~\ref{sec:construction}, here the positivity condition on
$a_{i,k}$ cannot be relaxed, because the transformation of sequence
space coordinates described in \cite{Krug_2024} cannot be applied
simultaneously to the different components of the phenotype vector.

\subsection{Basins of attraction}

It was observed in \cite{Krug_2024} that, for landscapes that satisfy the
AP, 
the subset and the superset of a peak genotype combined must be
contained in the
\textit{adaptive basin of attraction} (ABoA) of the peak, which is defined as the
number of sequences from which the peak is accessible via
fitness-monotonic paths. Since the number of elements
in the subset and the superset of a binary sequence $\sigma$ can be
written as $2^{|\sigma|} - 1$ and $2^{L - |\sigma|} - 1$ respectively,
the size of the ABoA, $S_\sigma$ of $\sigma$ obeys \cite{Krug_2024}
\begin{equation}
  \label{eq:BoA}
	S_\sigma \ge 2^{|\sigma|} + 2^{L - |\sigma|} - 2.
\end{equation}
The existence of a lower bound on the size of ABoA's that grows exponentially with the size of the genotype space is striking
and unexpected.  The bound is from below since the peak could additionally be accessible from other
sequences not in its sub- or supersets, and exploratory numerical simulations indicate that typical basins are often much larger
than suggested by \eqref{eq:BoA} \cite{Krug_2024}. Note that in the
absence of any contraints, the minimal size of an ABoA is equal to the
number of neighboring genotypes of a peak, which equals $L$ in the binary case.

The concept of \textit{adaptive} basins of attraction differs
importantly from the more commonly used notion of \textit{gradient}
basins, which contain those genotypes that are connected to the peak
by the (unique) path of steepest ascent \cite{Stadler2010}. Gradient basins are
disjoint and their union exhausts the state space, whereas adaptive
basins of different peaks generally overlap. While a few previous
studies have addressed the properties of 
gradient basins for NK fitness landscapes
\cite{Weinberger_1991,Tomassini2008,Franke2012}, little appears to be known about
adaptive basins.
A simple estimate\footnote{D. Oros
  and J. Krug, unpublished.} for the typical size of ABoAs for
the HoC model can be derived from the scenario of accessibility
percolation
that was explained in Sec.~\ref{sec:accessibility}. For large $L$,
the fitness of a peak is close to 1 in the uniform fitness scale, and
therefore the fraction of genotypes at scaled distance $\delta =
d/L$ from the peak that are connected to the peak through at least one
accessible path is equal to $1-\beta_c(\delta) > 0$. Thus typical ABoAs
contain a positive and principally computable fraction of all
genotypes. This suggests that the recent reports of extensive
adaptive basins in experimental high-throughput studies \cite{Papkou_2023,Westmann2024} may in fact
reflect a behavior that is generic rather than exceptional.

\section{Conclusion}

These notes have introduced probabilistic
models of biological fitness landscapes,
and reported recent progress in understanding their structural features. While the models are
very similar, and sometimes even equivalent to the energy landscapes
of interest in the statistical mechanics of disordered
systems \cite{Kauffman_1987,Park_2020,Das2022}, the evolutionary perspective suggests novel research
questions and brings in novel concepts. These include, for example, the quantification of
accessibility through fitness-monotonic paths and adaptive
  basins, as well as the notion of universal
epistasis, which is closely linked to sub- and supermodularity of
set functions.

Let us highlight
two specific scenarios of evolutionary accessibility that
have been described in this text. First, in uncorrelated random landscapes,
accessible paths between distant genotypes emerge through a
percolation-like transition when the fitness difference between the
endpoints is increased \cite{Schmiegelt_2023}. Second, structured landscapes that are submodular display
a peculiar organization of accessible paths into large basins of
attraction \cite{Krug_2024}. Since submodularity was argued to arise readily in
fitness landscapes with an intermediate (linear) phenotype, this
suggests a possible explanation for the recent observation of
large-scale empirical fitness landscapes that display many peaks with large basins of attraction
\cite{Papkou_2023,Westmann2024}.

To conclude, 
we hope that these brief notes will help to motivate further work exploring cross-connections between
the diverse manifestations of high-dimensional random landscapes, in statistical physics, evolutionary biology and computer
science. In particular, the burgeoning field of machine learning appears to present a rich set of questions for which the
concepts described here may potentially prove to be useful
\cite{Bahri2020}. In this context, we note that composite maps of the form \cref{eq:fisher} (typically including multiple intermediate layers)
        can be realized by neural network models
        \cite{Manrubia2021,Bahri2020}. In \cite{Pokusaeva2019} such a network was used to infer an intermediate linear phenotype and a nonlinear phenotype-fitness map from a large
        scale empirical data set for an essential enzyme in yeast. The
        composite maps in deep learning models often comprise
        elements that are sigmoidal rather than convex or concave, and
        therefore the ideas described in Sections \ref{sec:epistasis}
        and \ref{sec:submodular} do not obviously generalize to this
        setting. It is known that sigmoidal phenotype-fitness maps acting on a
        multi-dimensional phenotype can give rise to rugged fitness
        landscapes \cite{Schenk2013}, but their accessibility
        properties remain to be explored.

\section*{Acknowledgements}
The work described here is a collaborative effort over many years. In
particular, Kristina Crona, Suman Das, Jasper Franke, Muhittin Mungan,
Daniel Oros and
Benjamin Schmiegelt contributed essential insights and
results. Special thanks are due to Peter M\"orters and Jorge Pe\~{n}a
for pointing us towards the relation to submodularity.  

% TODO: include author contributions
% \paragraph{Author contributions}
% This is optional. If desired, contributions should be succinctly described in a single short paragraph, using author initials.

% TODO: include funding information
\paragraph{Funding information}
This work has been supported by Deutsche Forschungsgemeinschaft (DFG) within TRR 341 \textit{Plant Ecological Genetics} and
CRC 1310 \textit{Predictability in Evolution}.

\begin{appendix}
\numberwithin{equation}{section}

\end{appendix}


% Provide your bibliography here. You have two options:

%%% FIRST OPTION
% Write your entries here directly, following the example below, including:
% Author(s), Title, Journal Ref. with year in parentheses at the end, followed by the DOI number.

%\begin{thebibliography}{99}
%\bibitem{1931_Bethe_ZP_71} H. A. Bethe, {\it Zur Theorie der Metalle. i. Eigenwerte und Eigenfunktionen der linearen Atomkette}, Zeit. f{\"u}r Phys. {\bf 71}, 205 (1931), \doi{10.1007\%2FBF01341708}.
%\bibitem{arXiv:1108.2700} P. Ginsparg, {\it It was twenty years ago today... }, \url{http://arxiv.org/abs/1108.2700}.
%\end{thebibliography}

% Use your bibtex library, formatted by the SciPost style file.
\bibliography{SciPost_Example_BiBTeX_File.bib}

%%%%%%%%%% END TODO: BIBLIOGRAPHY

\end{document}